\documentclass[prb,showpacs,twocolumn,floatfix,showkeys]{revtex4}
\usepackage{graphicx} 
\usepackage{dcolumn}  
\usepackage{bm}       
\usepackage{amsmath}
\usepackage{amsfonts}
\usepackage{amssymb}
\usepackage{times}
\usepackage{subfigure}  
\begin{document}

\title{Crystal structure and electronic states of tripotassium picene}

\author{P.L. de Andres}
\affiliation{Departamento de Nanoestructuras, superficies y recubrimientos,
Instituto de Ciencia de Materiales de Madrid (CSIC), Cantoblanco, 28049 Madrid,
Spain.}

\author{A. Guijarro}
\affiliation{Departamento de Qu\'{\i}mica Org\'anica and
Instituto Universitario de S\'{\i}ntesis Org\'anica, Universidad
de Alicante, San Vicente del Raspeig, 03690 Alicante, Spain.}

\author{J.A. Verg\'es}
\affiliation{Departamento de Teor\'{\i}a y Simulaci\'on de Materiales,
Instituto de Ciencia de Materiales de Madrid (CSIC), Cantoblanco, 28049 Madrid,
Spain.}
\email{jav@icmm.csic.es}

\date{October 28, 2010}

\begin{abstract}
The crystal structure of potassium doped picene with an exact stoichiometry
(K$_3$C$_{22}$H$_{14}$, K$_3$picene from here onwards)
has been theoretically determined within Density Functional
Theory allowing complete variational freedom of the crystal structure
parameters and the molecular atomic positions. A modified herringbone
lattice is obtained in which potassium atoms are intercalated between
two paired picene molecules displaying the two possible
orientations in the crystal.
Along the c-axis, organic molecules alternate with chains
formed by three potassium atoms. The electronic structure
of the doped material resembles 
pristine picene, except that now the bottom of the conduction band is occupied
by six electrons coming from the ionized K atoms (six per unit cell).
Wavefunctions remain based mainly on picene molecular orbitals
getting their dispersion from intralayer 
{\it edge to face} CH/$\pi$ bonding, while eigenenergies 
have been modified by the change in the electrostatic potential.
The small dispersion along the $c$ axis is assigned to small
H-H overlap. From the calculated electronic density of states we
expect metallic behavior for potassium doped picene.
\end{abstract}



\maketitle

\section{Introduction}

Recently, Mitsuhashi et al. (Ref. \onlinecite{mitsu2010}) have shown
the existence of superconductivity at relatively high temperatures
for alkali-metal-doped picene. In particular, K$_{3.3}$picene shows a
critical temperature of $T_{c}=18$ K. This work opens the door to a new class
of materials that promise improved $T_c$ values. While the structure
of the organic picene crystal is well known since 1985\cite{de1985},
the best possible structural determination using synchroton radiation
on K doped samples has been indecisive about the actual
positions of interstitial atoms \cite{supp}.
No further advance in the understanding of the
superconductivity is possible until a successful detailed structural
determination exists. Once the atomic structure is known, both the
electronic states and the phonons can be quantitatively studied
and the usual ideas to understand superconductivity followed.
A preliminary theoretical determination of the crystal structure
of pristine and potassium doped picene 
(K$_{3}$picene) has been reported by Kosugi et al. \cite{kos2009}
by a first-principles calculation which adheres strictly to
experimentally determined crystal parameters for the
unit cell vectors length (see also comment in Ref. \onlinecite {metastable}).
Kosugi et al. find that
potassium doped picene molecules 
show an edge-to face intralayer ordering but stay almost
perpendicular to the ab plane, losing
the conventional herringbone interlayer stacking
(see Fig. 4a of the referred paper).
In our work, to circumvent 
the difficulties found in the experimental determination of the
structure of K$_{2.9}$picene (small and disordered samples, possible
existence of more than one structural phase, etc. \cite{mitsu2010})
we have adopted a more general strategy to the structural search of
the crystalline structure of highly doped picene;
we rely on
the best available {\it ab initio} methods based in Density
Functional Theory (DFT)
to get a stable structure that provides at least a local
minimum of the free energy of a precise stoichiometry.
We study the case of three potassium atoms
per picene molecule because it is the one showing the higher critical
temperature. Therefore, in our calculation atomic positions and parameters
defining the unit cell (including the symmetry group) have
been fully optimized to reach a total energy minimum configuration subject
to minimum forces and stresses.
The resulting crystalline structure
shows enough internal consistency and symmetry
to be theoretically appealing. We remark that the
calculated structure
differs significatively from two different structures
published for a smaller K doping of a similar material (pentacene crystal)
\cite{han2006,cra2009}.

\begin{figure}
\begin{center}
\subfigure{
\includegraphics[width=\columnwidth]{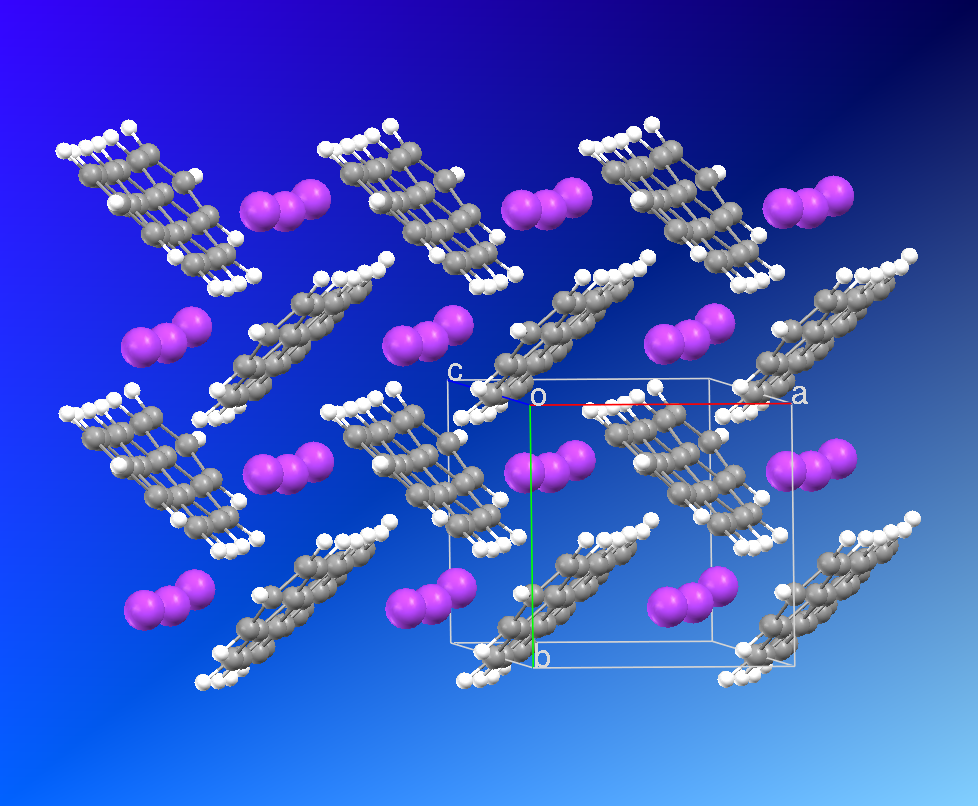}
}
\subfigure{
\includegraphics[width=\columnwidth]{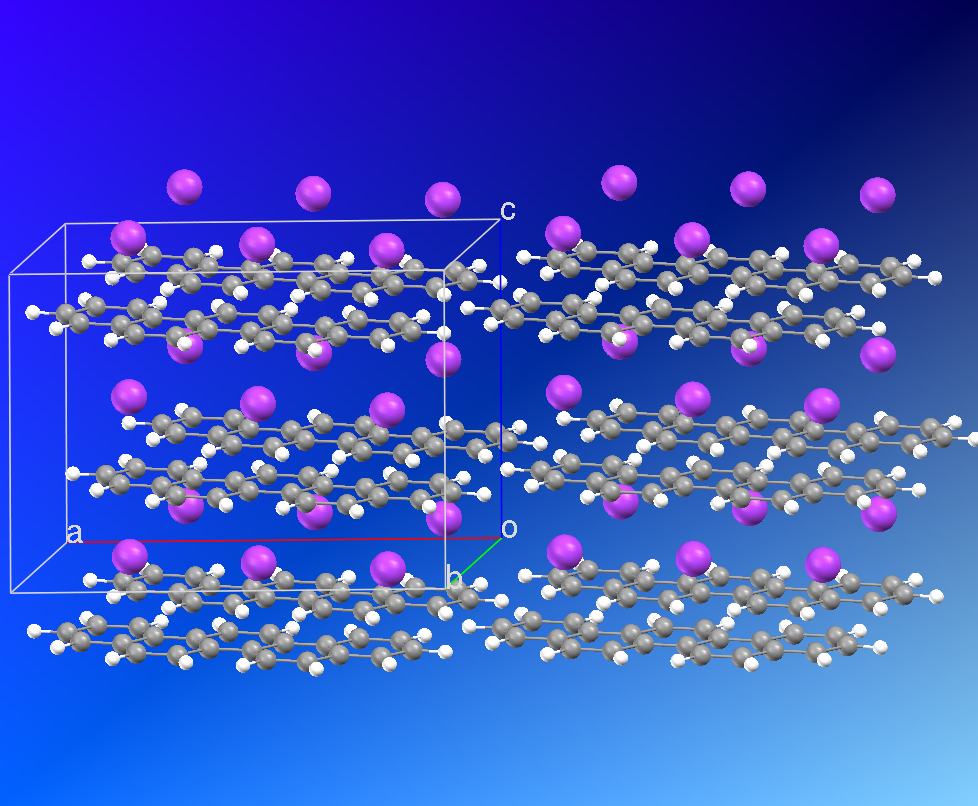}
}
\caption{(Color online)
Upper panel shows the
optimized atomic structure of K3picene as obtained with CASTEP for a
completely free lattice structure. The P2$_1$ spatial
symmetry of
pristine picene is conserved. Notice that although only one atomic layer
is displayed, the positions of the next-layer picene molecules are
on top of potassium atoms.
Lower panel shows an alternative structure obtained starting the
geometrical optimization from a laminar arrangement
of picene molecules intercalated by K atoms in a global
configuration based on intercalated graphite.
}
\label{estructura}
\end{center}
\end{figure}

The rest of the paper is organized as follows: Section II gives details
for the computational procedure followed in this work.
Section III presents our optimized crystalline
structure and the corresponding
electronic band structure and density of states. 
Finally, Section IV summarizes our conclusions.

\section{Numerical procedure}

A full experimental determination of the doped
picene crystal structure based in X-rays diffraction data
has not been possible so far because of the large background
and the small number of useful measured peaks. 
However, a basic Le Bail determination of the structure has
been performed to obtain approximate values for the unit
cell parameters\cite{supp}, yielding values quite similar to the
ones derived for pristine picene.
Therefore, we propose to complement the presently
limited experimental information with a pure numerical
procedure based in Density Functional Theory
to define realistic candidates for the actual crystalline structure.
One important advantage is that
Polycyclic Aromatic Hydrocarbons (PAH) show a robust geometry even in the
presence of alkali metals and translate and rotate almost rigidly during
the total energy minimization. However, the position of the
dopant atoms is a complex problem
not fully understood, as shown by previous theoretical
studies concluding different alternative models\cite{varios}.
In order to get a plausible crystalline structure for K$_3$picene,
we have minimized the free energy per unit cell using a DFT
scheme at the Local Density Approximation (LDA) level.
Both atomic positions and cell parameters
($a, b, c, \alpha, \beta, \gamma$) are free to change to minimize
forces on atoms and stresses on the unit cell get.
Some more technical parameters of the computation follow.
We use norm-conserving pseudopotentials\cite{opium},
an energy cutoff of 660 eV,
and Monkhrost-Pack\cite{monk} 
meshes of $4 \times 6 \times 3$ or $5 \times 5 \times 3$
depending on the unit cell being under consideration.
Actual calculations have been performed with the
CASTEP program allowing for spin polarization of
different bands of the doped crystal\cite{Payne,Accelrys}.
To describe the exchange and correlation potential
we use Kohn and Sham's local density approach
as computed by Ceperley and Alder\cite{Kohn,Ceperley}.
Maximum forces and stresses on the system are converged to
target thresholds of $0.1$ eV/{\AA} and $0.1$ GPa.

\begin{figure}
\begin{center}
\includegraphics[width=\columnwidth]{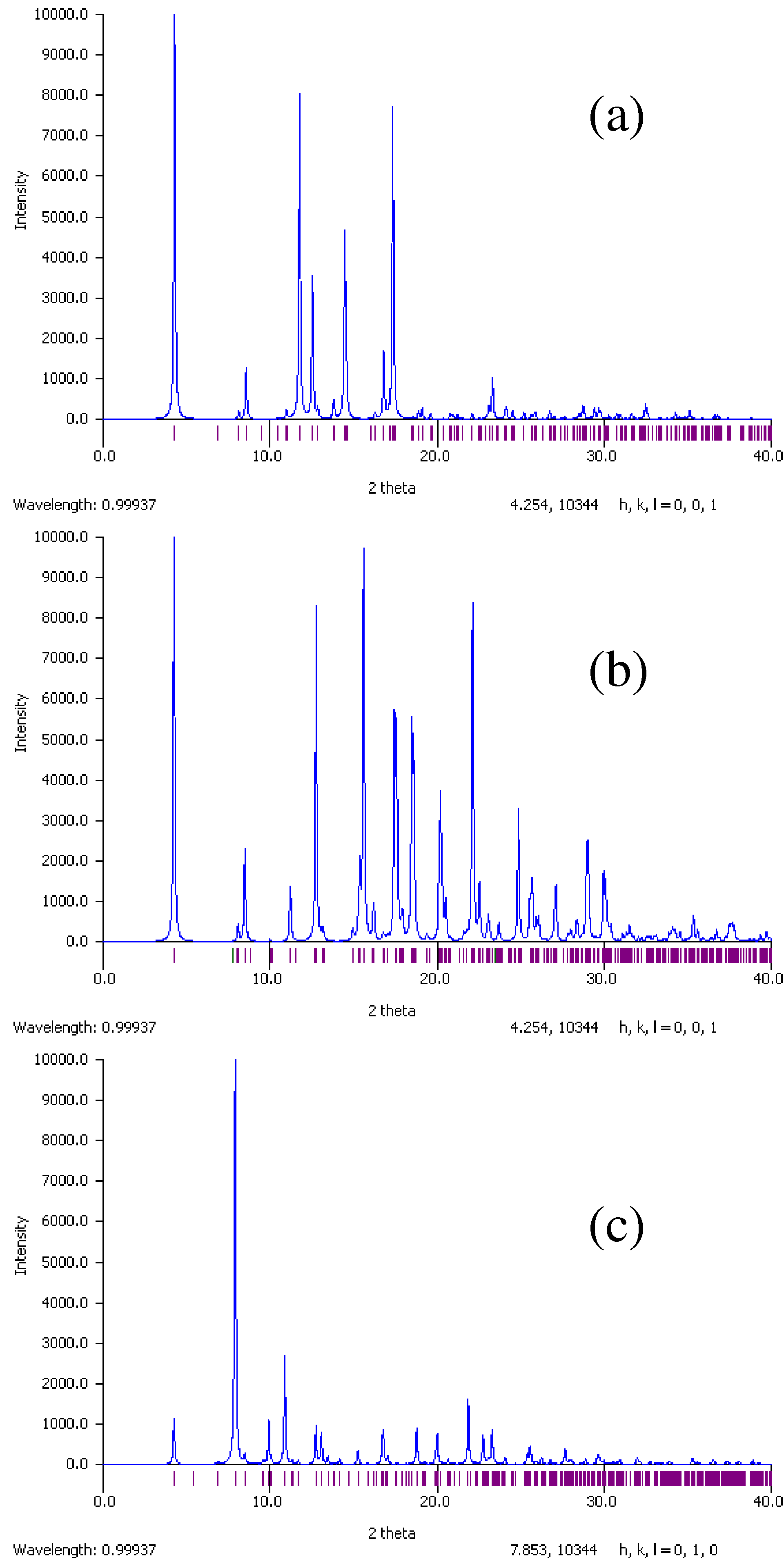}
\caption{(Color online)
Calculated powder X-ray diffraction pattern of the optimised 
theoretical structure of pristine picene (upper panel), 
K doped picene in the optimized herringbone structure (medium panel),
and K doped picene in the alternative laminar structure (lower panel).
Intensity is given in arbitrary units while diffraction angle is measured
in degrees.}
\label{RayosX}
\end{center}
\end{figure}

The whole procedure has been applied to pure picene as an
overall check. The optimization was started using the complete
structural determination provided by the Cambridge Crystallographic Data
Centre (CSD\_CIF\_ZZZOYC01). Cell parameters ($a, b, c, \beta$) 
change from
(8.480 \AA\ , 6.154 \AA\ , 13.515 \AA\ , 90.46$^{\circ}$) to
(8.373 \AA\ , 6.043 \AA\ , 13.390 \AA\ , 90.18$^{\circ}$)
improving the cell total energy in merely $0.12$ eV.
The unit cell volume shrinks about 4\%, but
the change in the molecule structure is imperceptible.
Therefore, the test is 
quite satisfactory and we assume that the differences
between experimental and computational values
provide an estimate for the errors
involved in our structural search.

\section{Results and Discussion}

\begin{figure}
\includegraphics[width=\columnwidth]{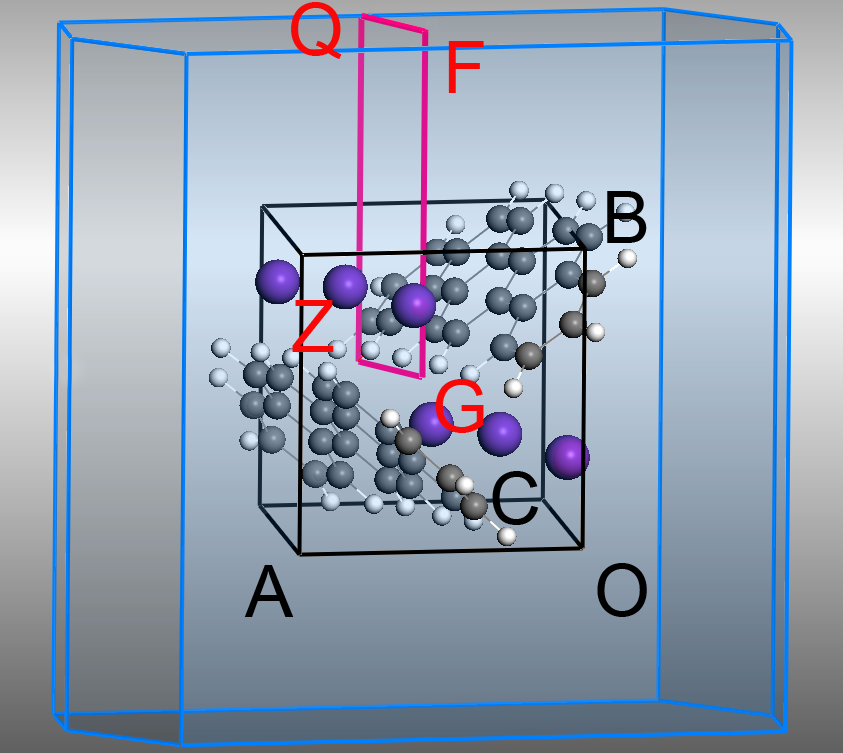}
\caption{(Color online)
Brillouin Zone obtained for K doped picene showing the crystalline
unit cell and molecules therein. The $\Gamma$FQZ$\Gamma$
path followed to calculate the band structure is shown in red.
}
\label{BZ}
\end{figure}

\subsection{Crystal Structure}

\begin{figure*}
\includegraphics[width=\textwidth]{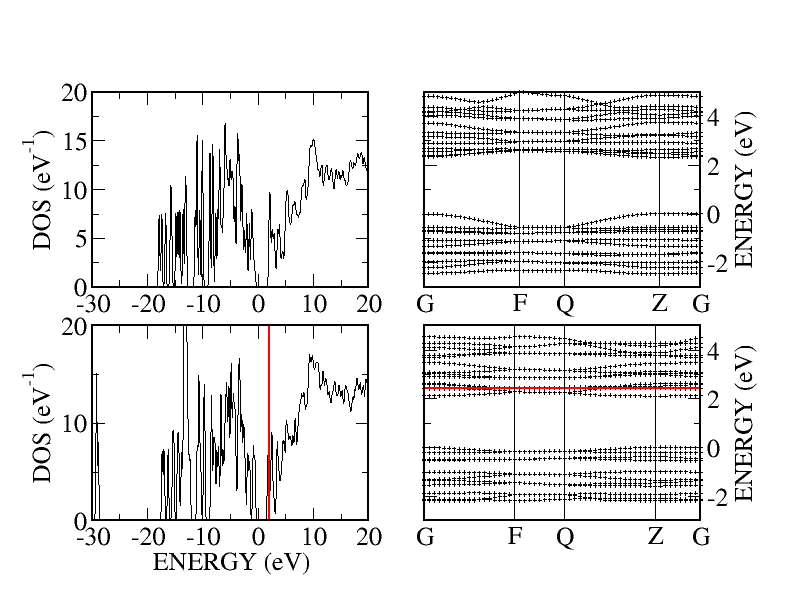}
\caption{(Color online)
Density of states and bands around the Fermi level for pristine
picene as described by the LDA DFT scheme use in this work (upper panels)
and the corresponding one after doping with potassium and relaxing
the herringbone original structure (lower panels).
The top of the valence band of both compounds has been aligned to
facilitate the comparison. Solid (red) lines indicate
the Fermi level position for the metallic system.
}
\label{DOS}
\end{figure*}

Previous work on K-doped pentacene\cite{han2006} and on
K$_3$picene\cite{kos2009} has proven that when large enough amounts
of alkali atoms are introduced in a herringbone organic
structure, the preferred positions for the dopants is intralayer, 
i.e., in between planes defined by PAH molecules. The other possibility,
K atoms in the planes between PAH layers is energetically less favorable
but it has also been considered for K$_1$pentacene\cite{cra2009}. Since we are
mainly interested in the case of very high doping (K$_3$picene) because it
shows promising results as a new superconducting material,
we have assumed from the beginning that K occupies sites between
picene molecules. As a starting crystal structure, we use pristine
crystalline picene with K atoms in the centroids of
next-nearest-neighbors parallel picene molecules\cite{centroides}.
Subsequently, the system is allowed to freely evolve under
atomic forces obtained from a series of fully converged DFT
calculations to drive the molecules to equilibrium positions. 
We have performed three sets of different calculations: 
(i) we froze the unit cell parameters to the values 
obtained from a Le Bail X-ray analysis,
(ii) values similar to
the ones deduced from a Le Bail X-ray analysis, but relaxed 
by less than 10\% in order to diminish the structural
stress, and
(iii) fully optimized values for the unit cell lengths and angles.
The optimum configuration for the first case allows us to
minimize the maximum component of forces on any atom
below $F_{max}=0.03$ eV/{\AA}, but the maximum component
of the tensor stress is $S_{max}=5$ GPa, and the total energy
is above the global minimum found later by $1.42$ eV. 
The second case allows us to decrease $S_{max}=2$ GPa,
but the total energy is still above the best configuration by
$0.97$ eV. 
Finally, the fully optimized structure
($a = 7.359 $\AA\,
$b = 7.361 $\AA\,
$c = 14.018 $\AA\,
$\alpha = 90^{\circ}$
$\beta = 105.71^{\circ}$ and,
$\gamma = 90^{\circ}$ )
allows us to decrease the maximum force and stress below
$0.1$ eV/{\AA} and $0.1$ GPa respectively.
The LDA cell volume increases from 669 \AA$^3$ for
clean picene, to 
730 \AA$^3$ after
doping with potassium. Even after a full optimization
the system belongs to the space group $P2_{1}$ and
remains monoclinic within a $\pm 0.01$ {\AA} error
in the atomic positions. Also $a=b$ at the precision
of our computation. A more detailed analysis shows that picene molecules
acquire a small regular curvature that contrasts with the structure
of the free molecule which is completely planar.
Fig. \ref{estructura}a shows how K atoms are intercalated
between the picene molecules that form a layer. The high-symmetry of
the arrangement is clear. Precise structural information is
given in a supplementary {\it cif}-like file \cite{epaps}.

\begin{figure}
\includegraphics[width=\columnwidth]{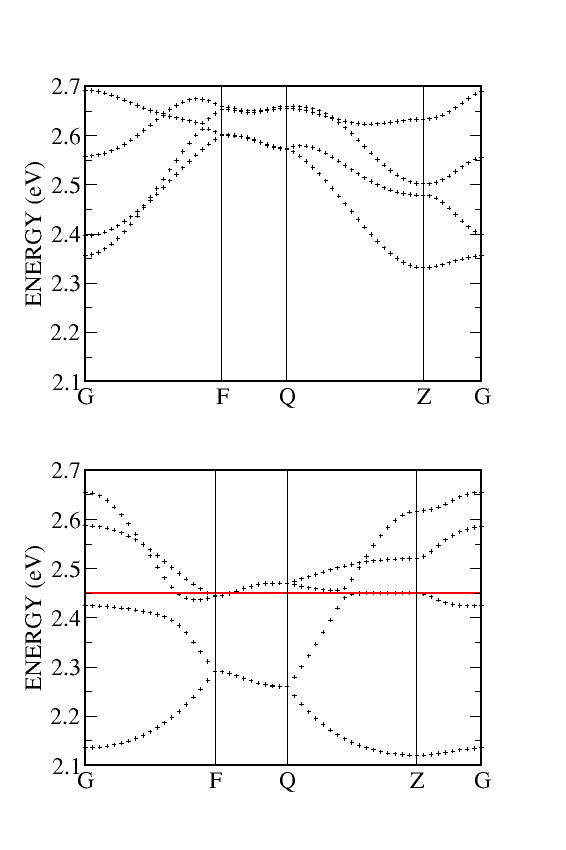}
\caption{(Color online)
Detail of the conduction bands occupied by K doping. Upper panel shows
the bands of pristine picene and lower panel the bands changed by the
presence of potassium.
}
\label{detalle}
\end{figure}

To understand the effect of different starting atomic configurations
on the final equilibrium structure, we have looked at a laminar structure
resembling intercalated graphite. The unit cell is made with two picenes
and six potassium atoms too, but now the picene
molecules are parallel. Potassium goes to centroids between corresponding
hexagonal rings. Starting from this configuration, DFT forces lead
to a locally stable structure in which picene molecules form a global plane
whereas intercalated potassium completes a structure resembling
the well known KC$_8$. The total energy of this structure is more
than two eV per cell higher than the previous herringbone one. Consequently,
from a strict energetic point of view, it is clearly inferior.
Fig. \ref{estructura}b shematically shows this laminar structure.  
Again, precise structural information is given in a supplementary
{\it cif}-like file \cite{epaps}.

\begin{figure}
\includegraphics[width=\columnwidth]{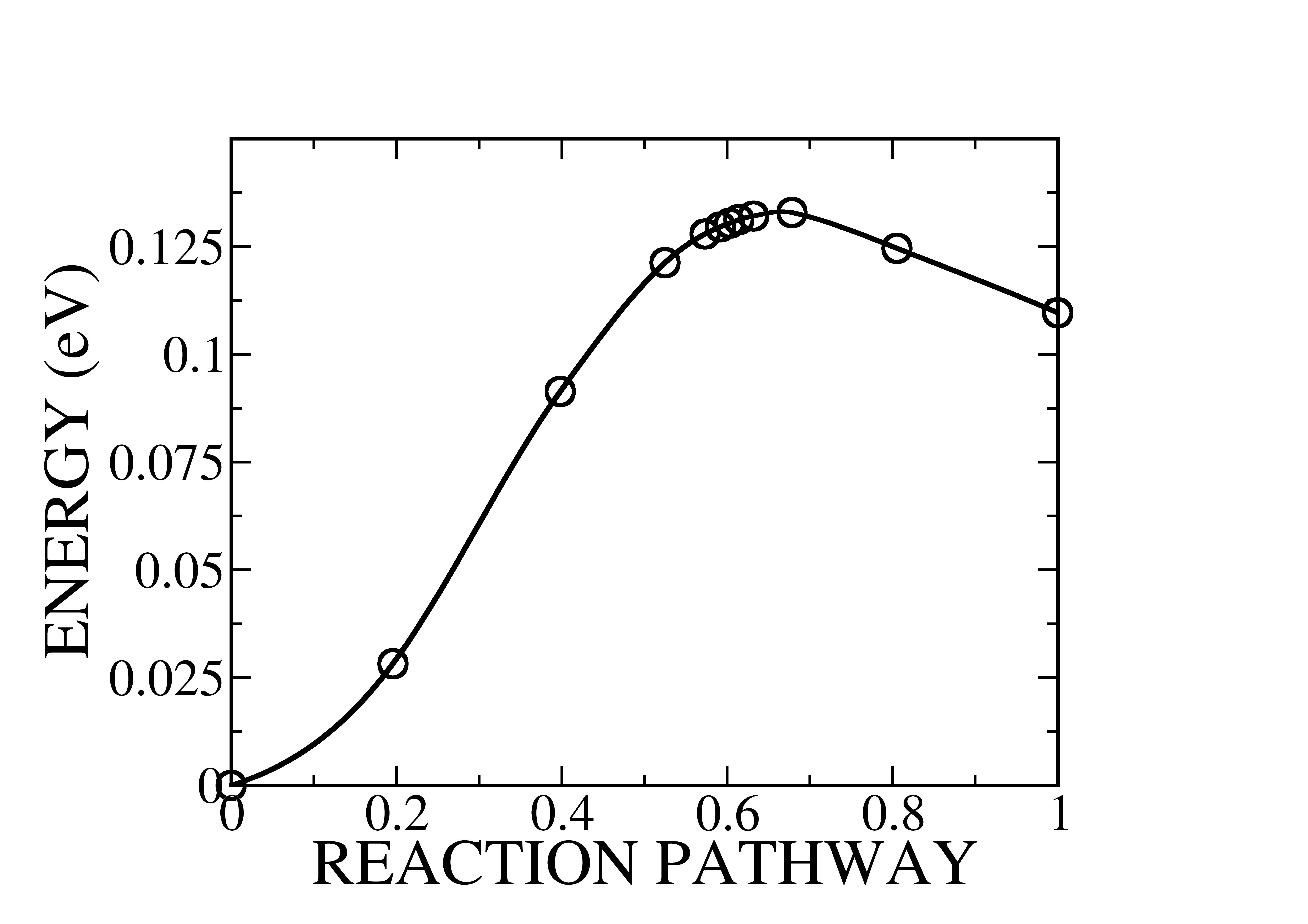} 
\subfigure{
\includegraphics[width=\columnwidth]{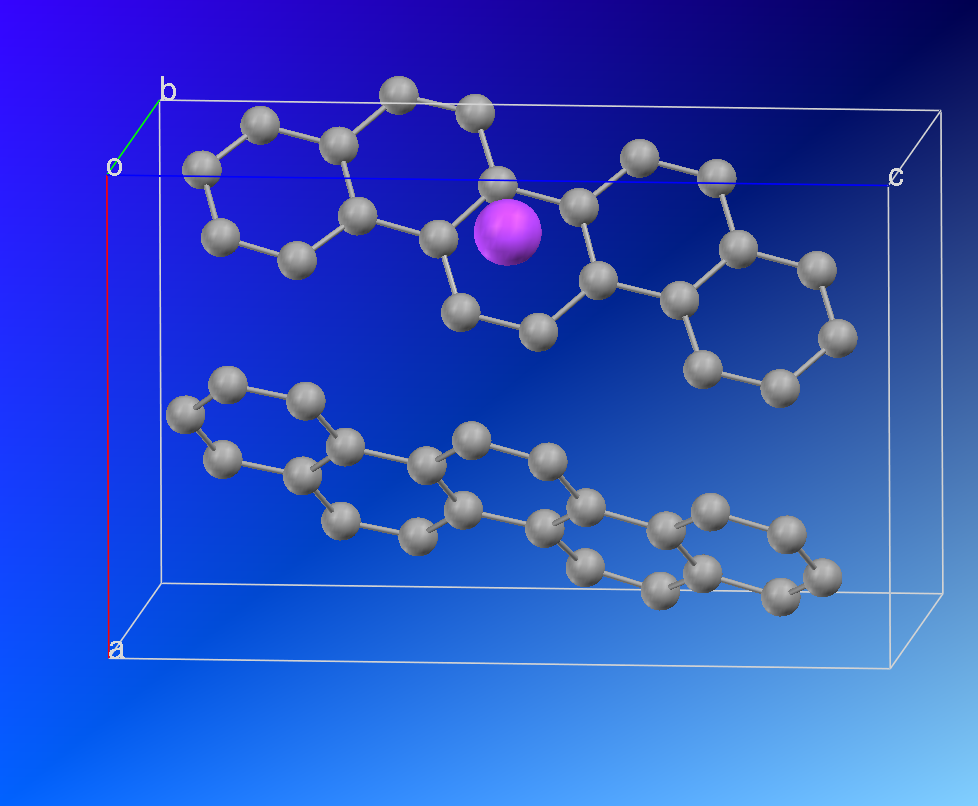} }
\subfigure{
\includegraphics[width=\columnwidth]{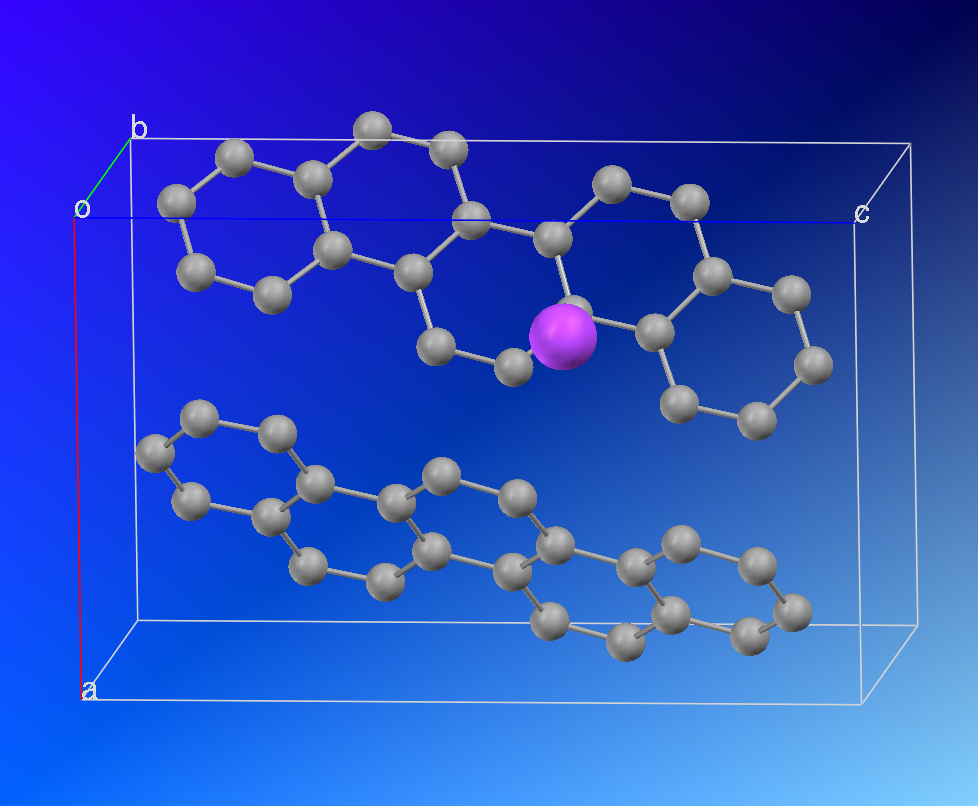} }
\caption{(Color online)
Diffusion barrier (eV) for one K atom diffusing
inside the frozen clean picene unit cell.
Configurations for the initial
and final states are shown in the upper and
lower panels, respectively
(hydrogens have been 
removed from the picture for the sake of clarity).
}
\label{barrier1K}
\end{figure}

The ultimate justification of a crystal structure should come
from experiment, typically from X-ray diffraction. Mitsuhashi et al.
\cite{mitsu2010} provide 
the X-ray diffraction pattern of a K$_{2.9}$picene sample obtained with
synchroton radiation of $\lambda = 0.99937 $\AA\
in the supplementary information of their paper. We have used the
mercury program\cite{mercury} to calculate a powder pattern for any
well defined crystal structure. Our results are shown in
Fig. \ref{RayosX}.
First, we have checked that both the original
picene crystal and the one refined by DFT at the LDA level provide
spectra similar to the one given in Ref. \onlinecite{mitsu2010}.
Second, we have used the same program to get the pattern
for our predicted K$_3$picene structure (see Fig. \ref{RayosX}).
It can be observed that the small angle part of the spectrum
of doped picene resembles the one obtained for pristine picene
in the herringbone structure. This is
easily explained since the arrangement of picene molecules is
still close to the original one. Until here there is some agreement
with experiment since Mitsuhashi et al. comment that X-ray diffraction
pattern of K$_{2.9}$picene shows a good correspondence with that
of pristine picene (Figs. S7a and S7b of Ref. \onlinecite{mitsu2010}).
Nevertheless our well ordered array of K atoms produces important
peaks at larger angles that have not been experimentally observed.
We believe that K disorder, phase mixing, etc. prevent a better
experimental
determination of dopant positions and demand more research
in the future.

\begin{figure}
\includegraphics[width=\columnwidth]{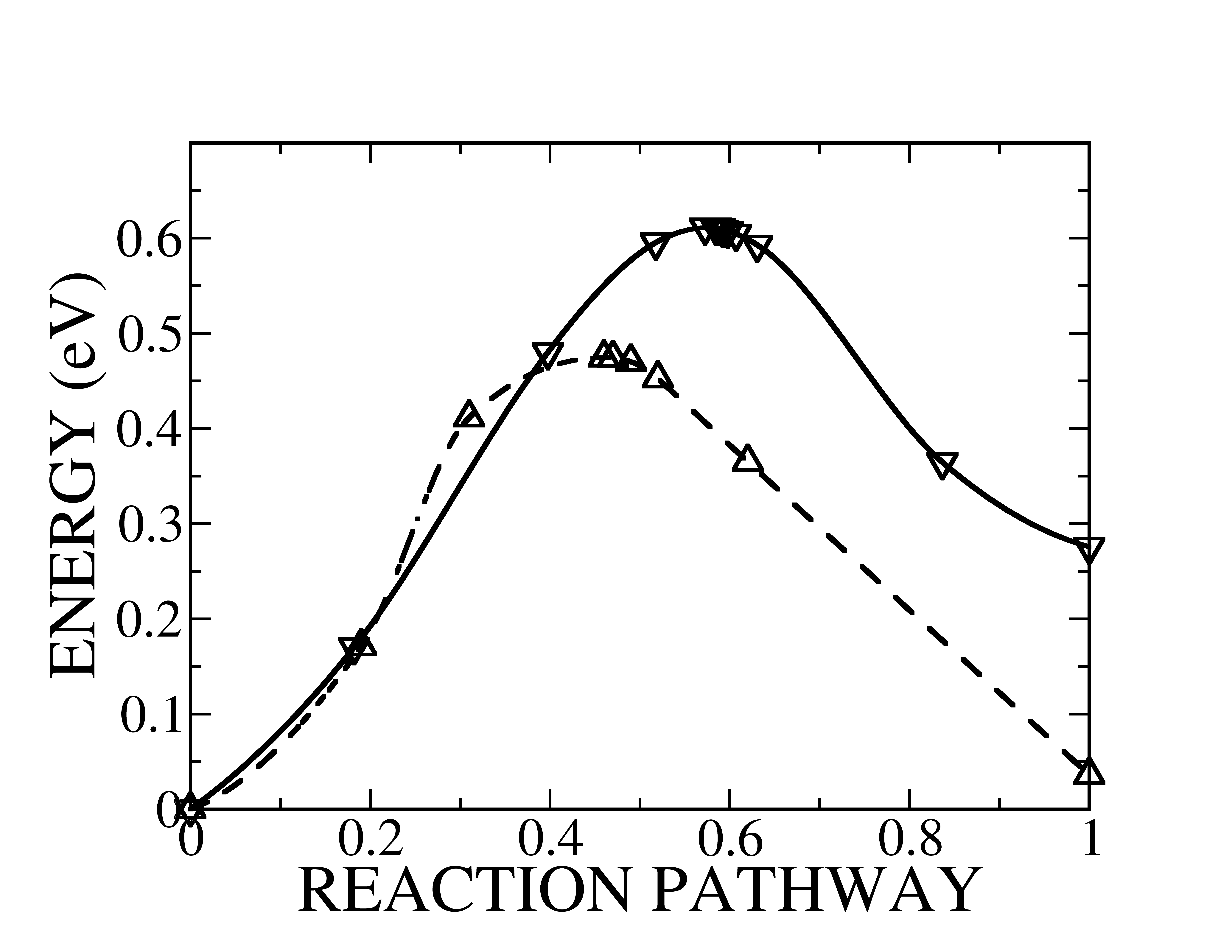} 
\subfigure{
\includegraphics[width=\columnwidth]{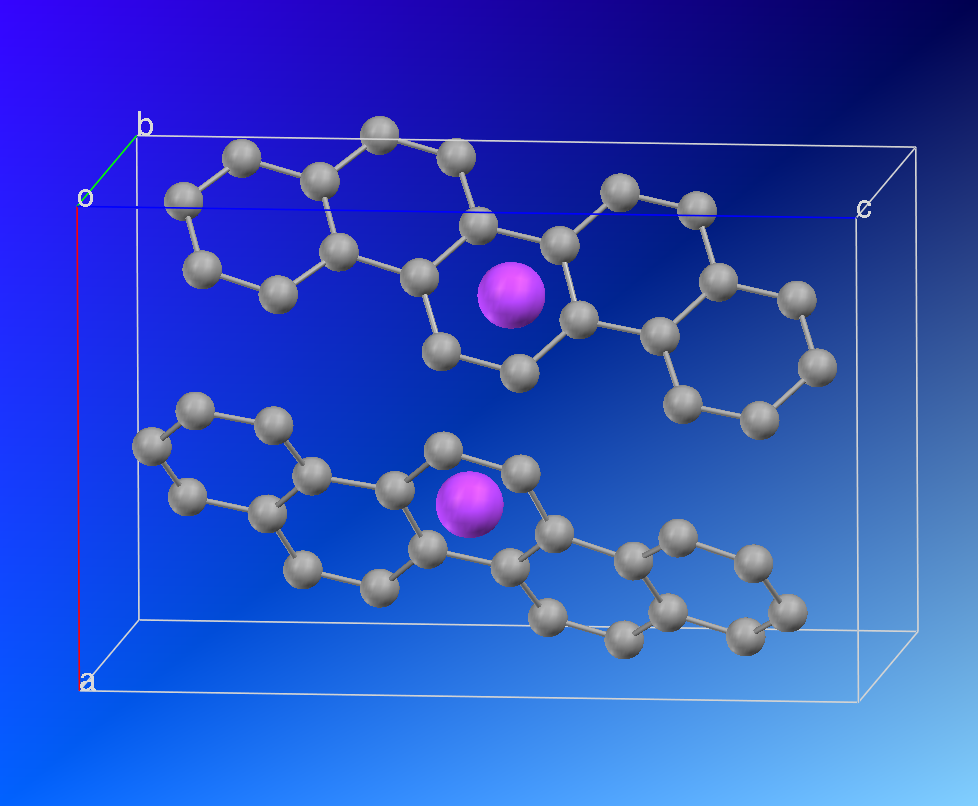} }
\subfigure{
\includegraphics[width=\columnwidth]{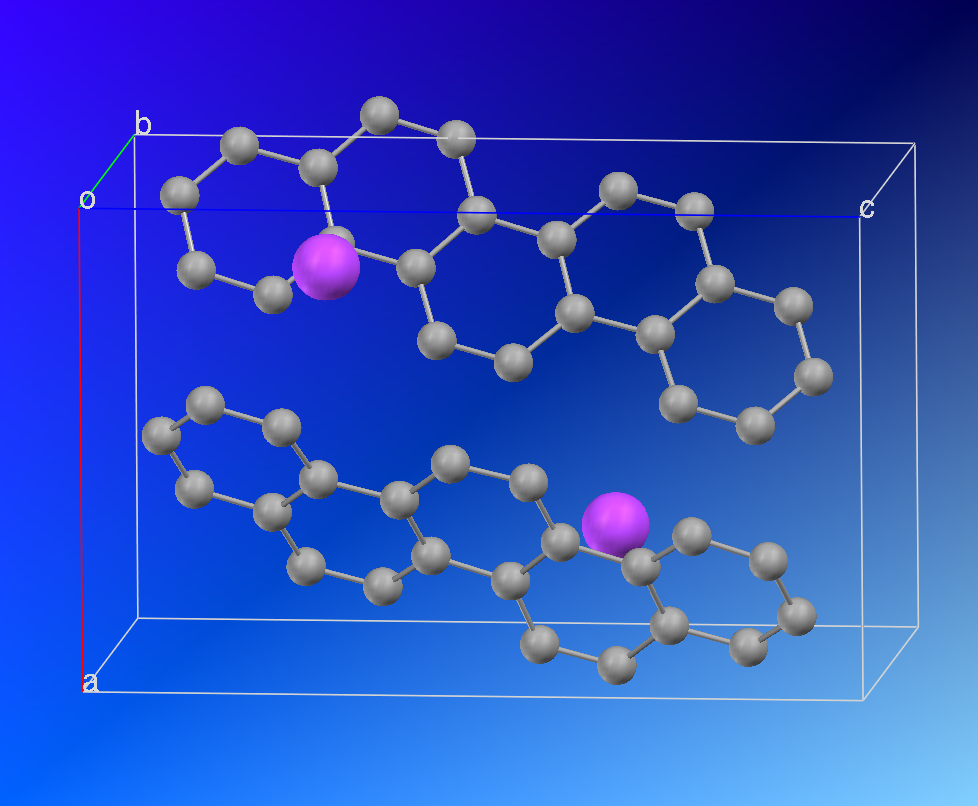} }
\caption{(Color online)
Diffusion barriers (eV) for two K atoms 
moving in the same direction 
(triangles pointing down, continuos lines)
and in the opposite direction
(triangles pointing up, dashed line). 
Configurations for the initial
and final states are shown in the upper and
lower panels, respectively.
}
\label{barrier2K}
\end{figure}

On the other hand, calculated powder pattern spectra allow us
to discard alternative laminar stuctures because they
systematically show the extintion of
the (0,0,1) low angle peak. It is easy to check
that the spectrum of potassium intercalated graphite
(KC$_8$) shows a similar property.

\subsection{Electronic Structure}

The motivation behind the structural work is to 
better understand the electronic properties of the material.
We compute the electronic density of states and the band structure
for the optimized configuration using standard
procedures in DFT.
Fig. \ref{BZ} gives some details of the Brillouin Zone form and
orientation relative to the crystal lattice.
Also the path used to calculate the band structure is shown.
Fig. \ref{DOS} gives results corresponding to the LDA clean picene
crystal (upper panels) and doped K$_3$picene in the LDA herringbone
crystalline structure (lower panels). The most significant feature of
picene electronic structure is the opening of a wide gap (about 2.5 eV
in our approximation) separating occupied (valence) states from
empty (conductance) states. The upper right panel shows the bands that
are close to the chemical potential. This result is in accordance with
the insulating character of picene although the gap is
underestimated in the local density approximation (LDA). 
After adding six potassium atoms per cell and assuming that the
theoretically proposed crystalline structure is correct, the material
presents the electronic structure shown in the lower panels
of Fig. \ref{DOS}. Both valence and conduction bands are globally similar to
the previous ones except for two new peaks appearing in the valence band,
the first just above -30 eV and the second around -13 eV. They
correspond to the fully occupied 3s and 3p states coming from
potassium atoms. The valence electrons of K atoms have been transferred to
the conduction band as the Fermi level position indicates (we have 
checked that the integral below the occupied part of the conduction band
is precisely six). Consequently, the potassium 4s states lie well above
in the conduction band (but are not shown in the Figure) and are
completely empty. A closer look to the states that are occupied
by K electrons in doped picene is given by Fig. \ref{detalle}.
It is clear that the dispersion of the bands is larger in the
doped material and consequently, the separation to the valence
bands is smaller. 
Notice also that dispersion along FQ and Z$\Gamma$ directions
(both parallel to ${\bf k}_{3}$ axis ($c$* axis) is more or less
similar. Although the assignment of wavefunctions weight to
particular atoms or molecules is beyond the scope of this contribution,
we believe that extended states of the doped material are similar to
pristine picene wavefunctions but their energies are shifted by the new
electrostatic potential induced from the transference of electrons from
potassium to organic molecules. Therefore, assuming that 4s K orbitals
do not have a significant weight in the conduction band states,
the existence of several weak
CH/$\pi$ bonds \cite{CHpi,CHpibis}
among perpendicular picene molecules can explain
the existence of extended states. On the other hand,
the smaller H-H overlaps should provide the smaller delocalization
along the $c$ axis.

Once changes in the electronic structure of picene due to the
doping by K have been analysed, some conclusions can be
inferred.
First, the introduction of potassium gives rise to an ionic system
in which long range Coulomb forces contribute an important
percentage of the whole cohesive energy.
Second, the broadening of the molecular levels of
individual picene molecules likely comes from the original
edge to face interactions of any herringbone arrangement
plus the smaller interlayer coupling mediated by H-H overlap. Consequently,
the doped system presents extended states covering the whole crystal,
but showing a reduced dispersion along the ${\bf k}_{3}$ axis.
This characteristic implies a smaller electric
conductivity parallel to the $c$ axis in comparison with
conductivities in the $a$ and $b$ directions. Measurements
mentioned by Mitsuhashi et al. (Ref. \onlinecite{mitsu2010}) for
doped pentacene share this property although physical
grounds are quite different. In the last case, conductivity shows
a two-dimensional character because doping occurs in $a$-$b$ planes
allowing extended two-dimensional wavefunctions for the
alkali metals.

\subsection{\label{sec:lstqst}Intercalation of K}

Intercalation of K atoms inside the picene crystal is crucial
to modify some properties of interest, e.g. superconductivity.
In this subsection, we use our structural and electronic calculations
to compute diffusion barriers for K atoms
moving inside the unit cell. 
The starting point are 
local equilibrium adsorption sites
computed for a common unit cell that we choose
to be the experimental one for clean picene\cite{de1985};
this is a reasonable choice because of the low
amount of potassium involved in these calculations. 
Barriers are estimated 
from a Linear Synchronous Transit (LST)
transition state search, followed by a Quadratic Synchronous Transit
(QST) refinement\cite{Halgren97}.

The global energy minimum for a single K in the picene
unit cell is close to the central part of the channel defined
by four picene molecules (Figs. \ref{estructura}a and \ref{barrier1K}).
We compute the diffusion barrier to move the K atom
along the channel (the displacement introduces new forces
making the atom to get closer to one
of the picene molecules by 0.6 {\AA}). 
The energy of reaction when moving between these two
local adsorption sites is 0.11 eV and the highest barrier
(from the initial position to the transition state) is 0.13 eV
(Table \ref{table1}). 
This barrier is small enough as to predict a large diffusivity of
K inside the picene crystal: we estimate this value from
$W e^{-B/k_{B} T}$, where $W$ is a typical attempt
frequency that we take as the frequency of a plausible
phonon (e.g., $1$ THz), B is the diffusion barrier
for this particular hop from site I to site F,
$T$ is temperature and $k_{B}$ is the
Boltzmann constant.
This simple estimate shows that the potassium atom
is making about $10^{9}$ hops per second at room temperature, 
making macroscopical diffusion in a reasonable
amount of time possible.

For two K simultaneously diffusing in the unit cell many more
configurations need to be explored making an extremely complex
energy landscape. We show in Fig. \ref{barrier2K}
two representative cases: starting out from the global minimum
configuration where both K atoms stay near the center of the
unit cell, both atoms diffuse either along the same or opposite 
directions (mostly, both along the $+c$, or the $+c$ and $-c$ axis
respectively). The lower barrier corresponds to the
K atoms moving away from each other:
0.47 and 0.44 eV from the initial and final states respectively. 
For the {\em concerted} diffusion, barriers are 
0.61 and 0.33 eV respectively, and the initial and
final states have a larger energy difference:
0.28 eV  (Table \ref{table1}).

\begin{table}
\caption{
Barriers from the initial ($B_{I}$) and final ($B_{F}$)
states, and energy of reaction ($\Delta E=E_{F}-E_{I}$) for
one and two K atoms diffusing in the unit cell 
(primarily along the c direction). For two K atoms
two different cases have been considered:
a symmetric (antisymmetric) diffusion mode (S or A, respectively) 
where both atoms  move along the same (opposite) directions.
}
\label{table1}
\begin{tabular}{c|rrr}
\hline
System & B$_{I}$ & B$_{F}$ & $\Delta E$ \\
\hline
One K            & 0.13 & 0.02 & 0.11 \\
Two K (S)  & 0.61& 0.33 & 0.28\\
Two K (A) & 0.47 & 0.44 & 0.04 \\
\end{tabular}
\end{table}

\section{Concluding Remarks}

Systematic use of DFT allows the prediction of a plausible
crystalline structure for heavily doped picene (K$_3$picene).
Potassium is absorbed in the square prism space defined by four
equivalent picene molecules. The original herringbone structure formed
by the organic molecules is only weakly deformed. This justifies
the similarity of the X-ray powder spectra of the doped material and
the pristine picene. Nevertheless, our predicted structure do not
quantitatively reproduce the experimental spectrum, mainly because the
larger angle peaks originating in K are not present in experiment.
The metallic character of the doped material comes from the
occupation of picene wavefunctions due to the transfer of electrons
from potassium to the organic molecules.
The presence of K cations induce a modified electrostatic potential
that broadens the lower part of the conduction band without
appreciably changing the wavefunctions based on molecular
orbitals of picene.
Therefore the reason for the intralayer
dispersion of the bands
is the CH/ $\pi$ interaction responsible for the
herringbone PAH arrangement.
We should comment that this delocalization effect will be lost for
separated parallel picene molecules, providing an important argument
against alternative laminar structures.
In view of these results, we assume that
superconductivity might be triggered
by electron-phonon coupling {\em within} individual picene molecules.

\begin{acknowledgments}
The authors are grateful to E. Guti\'errez for helping us with
the interpretation of X-ray spectra.
We are also grateful to A. Hansson for providing us with the structural
data of the two phases of Kpentacene shown in Fig. 1 of
Ref. \onlinecite{han2006}.
Financial support by the spanish "Ministerio de Ciencia e Innovaci\'on
 MICINN" (grants CTQ2007-65218, CSD2007-00006,
MAT2008-1497, MODELICO and FIS2009-08744) is gratefully acknowledged.
We also acknowledge support from the DGUI
of the Comunidad de Madrid under the R\&D Program of activities
MODELICO-CM/S2009ESP-1691.

\end{acknowledgments}


\begin{thebibliography}{99}
\bibitem{centroides}
Initial positions for the six K atoms per cell
are obtained calculating the centroids of corresponding hexagonal
rings of parallel molecules. First, third and fifth rings between
parallel molecules separated by $a$ were chosen although the centroids
corresponding to the closer stacking ($b$) provide
almost the same initial positions.
\end{thebibliography}
\end{document}